\documentclass[12pt,preprint]{aastex}

\shorttitle{\indent \def Upflows in funnel-like legs of loops}
\shortauthors{Tian et al.}

\begin{document}

\title{Upflows in funnel-like legs of coronal magnetic loops}

\author{Hui Tian\altaffilmark{1,2}, Eckart Marsch\altaffilmark{2},
Werner Curdt\altaffilmark{2}, Jiansen He\altaffilmark{2}}

\altaffiltext{1}{School of earth and space sciences, Peking
University, 100871, Beijing, China; tianhui924@gmail.com}

\altaffiltext{2}{Max-Planck-Institut f\"ur Sonnensystemforschung,
37191, Katlenburg-Lindau, Germany}

\begin{abstract}
The prominent blue shifts of Ne~{\sc{viii}} associated with the
junctions of the magnetic network in the quiet Sun are still not
well understood. By comparing the coronal magnetic-field structures
as obtained by a potential-field reconstruction with the conspicuous
blue-shift patches on the dopplergram of Ne~{\sc{viii}} as observed
in an equatorial quiet-Sun region, we find that most of the regions
with significant upflow are associated with the funnel-like legs of
magnetic loops and co-spatial with increments of the line width.
These quasi-steady upflows can be regarded as the signatures of mass
supply to coronal loops. By using the square-root of the line
intensity as a proxy for the plasma density, the mass flux of the
upflow in each funnel can be estimated. We find that the mass flux
is anti-correlated with the funnel's expansion factor as determined
from the extrapolated magnetic field. One of the loop systems is
associated with a coronal bright point, which was observed by
several instruments and exhibited various morphologies in different
wavelengths and viewing directions. A remarkable agreement between
its magnetic structure and the associated EUV emission pattern was
found, suggesting an almost potential-field nature of the coronal
magnetic field. We also report the direct detection of a small-scale
siphon flow by both STEREO satellites. However, this transient
siphon flow occurred in a weak mixed-polarity-field region, which
was outside the adjacent magnetic funnel, and thus it is perhaps not
related to plasma upflow in the funnel. Based on these observations,
we suggest that at upper-TR (transition region) temperatures the
dominant flows in quiet-Sun coronal loops are long-lasting upflows
rather than siphon flows. We also discuss the implications of our
results for coronal heating and unresolved magnetic structures.
\end{abstract}


\section{Introduction}

Quasi-steady flows have been frequently observed everywhere in the
upper solar atmosphere, and the apparent steadiness of the global
flow pattern seems to indicate a systematic large-scale plasma
circulation in the solar corona and transition region (TR)
\citep{Foukal1978,Marsch2004,Marsch2008,Dammasch2008}. It is well
known, but not yet fully understood, that in the network of the
quiet Sun the ultraviolet emission lines formed in the TR are
redshifted by a few km/s
\citep[e.g.,][]{Doschek1976,Brekke1997,Chae1998,Curdt2008}. As the
temperature increases, the observed average Doppler shift turns from
a red into a blue shift in the upper TR
\citep{PeterJudge1999,Xia2004}. Recently, EIS \citep[EUV Imaging
Spectrometer,][]{Culhane2007} observations showed that coronal lines
revealed high outflow velocities on the order of 100~km/s in a
compact region (network boundary) of the quiet Sun \citep{Dere2007}.

The Ne~{\sc{viii}}~(770.4~{\AA}) line is formed in the upper TR and
lower corona and is on average blue shifted in coronal holes and the
quiet Sun \citep[e.g.,][]{Dammasch1999}. In coronal-hole
dopplergrams, sizable patches of blue shift were frequently reported
and usually interpreted as indicators of solar wind outflow
\citep{Hassler1999,Stucki2000,Wilhelm2000,XiaEtal2003,Aiouaz2005,Tu2005a}.
In the quiet Sun, significant blue shifts of Ne~{\sc{viii}} were
also found at the network junctions and considered to be possible
sources of the solar wind \citep{Hassler1999}. However, through a
combined analysis of and comparison between 3-dimensional (3-D)
magnetic-field structures and EUV observations, \cite{He2007} and
\cite{Tian2008a} found that most of the sites with Ne~{\sc{viii}}
blue shift were not located in open-field regions. Consequently,
they argued that these sites might not be sources of the solar wind.
Furthermore, \cite{Tian2008a} also noticed that there were some
loops revealing upflows in both legs, and some other loops with
upflow in one and downflow in the other leg.

Although the smaller coronal structures are not well resolved, many
attempts have already been made to understand these magnetic-field
structures above the photosphere. \cite{Gabriel1976} suggested the
first magnetic-network model, in which the TR emission originates
from magnetic funnels diverging with height and originating from the
underlying supergranular boundaries. This picture was modified by
\cite{Dowdy1986} who suggested that only a fraction of the network
flux shaped as a funnel opens into the corona, while the majority of
the network is occupied by a population of low-lying loops.
\cite{Peter2001} suggested that there may be two types of funnels,
namely the ones connected to the solar wind and those forming the
feet of large loops. Recently, based on the observed vector magnetic
field, \cite{Tsuneta2008} proposed that the height where the funnel
expands dramatically is higher in the quiet Sun than in coronal
holes. On the other hand, magnetic-field extrapolation is a common
method for the solar community to construct the coronal field and
study the magnetic coupling of different solar processes
\citep{WiegelmannNeukirch2002}. The linear force-free model as
proposed by \cite{Seehafer1978} has been successfully applied before
to study coronal holes and quiet-Sun regions
\citep{Wiegelmann2005,Tu2005a,Tu2005b,Marsch2006,He2007,Tian2007,Tian2008a}.

In this paper, we present new observational results and further investigate the
magnetic coupling and guidance of the plasma flows in corona and TR. The
results are discussed and interpreted in the context of coronal circulation.

\section{Ultraviolet observations and magnetic field extrapolation}

\begin{figure}
\centering {\includegraphics[width=\textwidth]{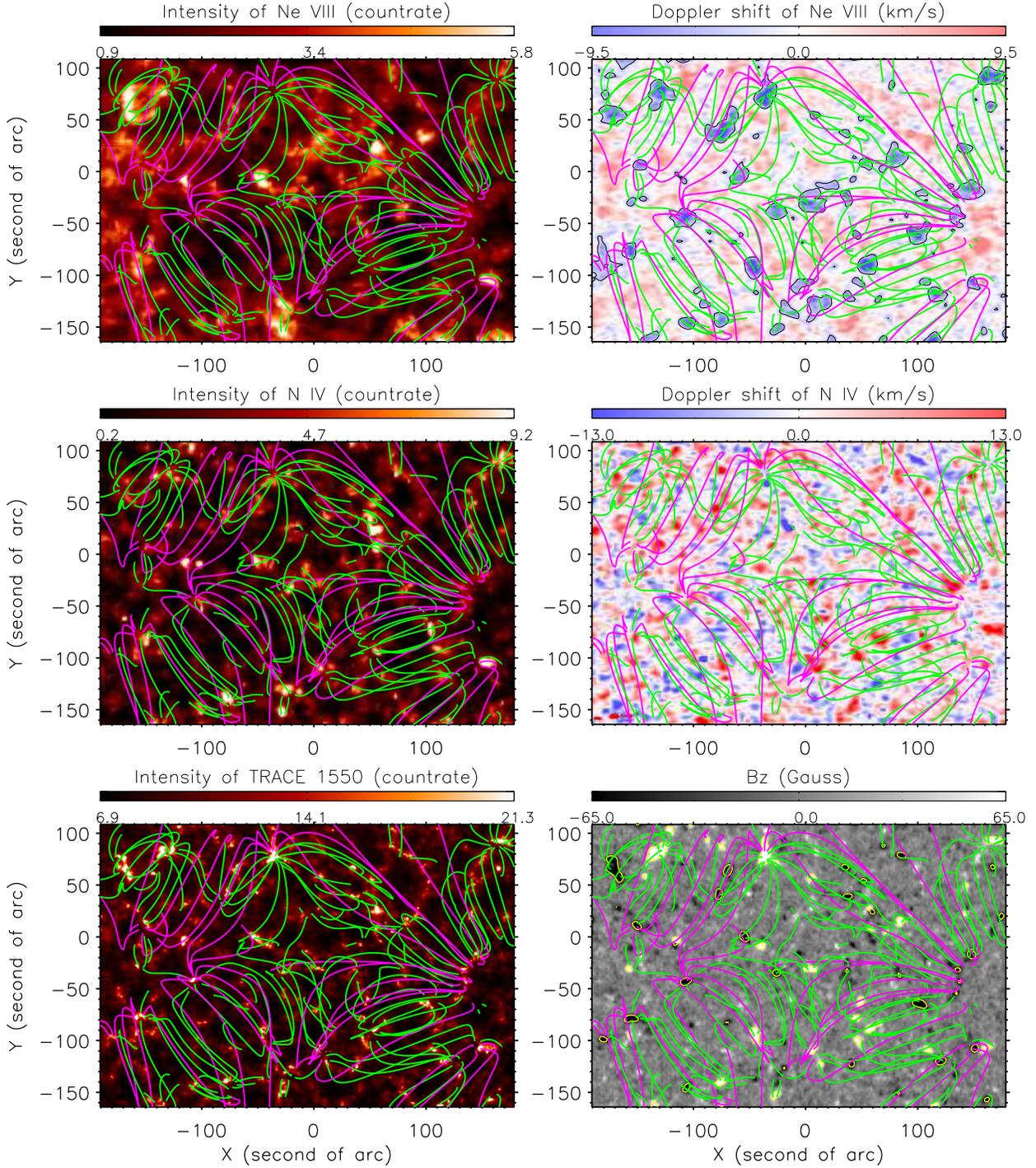}}
\caption{Projections onto the x-y-plane of the extrapolated magnetic
field lines, being superposed on the maps of the intensity and
Doppler shift of Ne~{\sc{viii}} (upper panels) and N~{\sc{iv}}
(middle panels), the intensity of TRACE 1550~\AA~(lower left), and
the photospheric longitudinal magnetic field strength (lower right).
Field lines reaching higher and lower than 40~Mm are plotted in
purple and green, respectively. Patches with significant blue shift
(larger than 3~km/s) are outlined by the filled contours on the
dopplergram of Ne~{\sc{viii}}. In the magnetogram, the strong-field
(larger than 40~gauss in field magnitude) regions are outlined in
yellow. } \label{fig.1}
\end{figure}

The SUMER (Solar Ultraviolet Measurements of Emitted Radiation)
\citep{Wilhelm1995,Lemaire1997} data analyzed here were acquired at
disk center from 13:24 to 16:00 UT on July 3, 2008. The slit 2
($1^{\prime\prime}\times300^{\prime\prime}$) was used to scan an
area with a size of about
$370^{\prime\prime}\times300^{\prime\prime}$ in this observation
sequence. Spectra of Ne~{\sc{viii}}~($\lambda$~770.4~{\AA}) and
several other TR lines including N~{\sc{iv}}~($\lambda$~765.1~{\AA})
were recorded on detector B with an exposure time of 60~s. The
standard SUMER procedures for correcting and calibrating the data
were applied. They include decompression, flat-field correction, and
corrections for geometrical distortion, local gain and dead time. By
applying a single-Gaussian fit to each spectrum, the intensity map
and dopplergram of Ne~{\sc{viii}} were obtained and presented in
Fig.~\ref{fig.1}.

During the SUMER observation, full-disk magnetograms with a 1-minute
cadence and 2\arcsec~pixel size were obtained by MDI (Michelson
Doppler Imager) \citep{Scherrer1995} onboard SOHO. We found that the
magnetogram was quite stable and showed almost no change during the
period of the SUMER observation. In order to increase the
signal-to-noise ratio, seven magnetograms observed from 15:00 to
15:06 UT were averaged. The coalignment between SUMER and MDI images
was achieved through a cross correlation between the radiance map of
N~{\sc{iv}} and the averaged magnetogram. The coaligned magnetogram
can be found in Fig.~\ref{fig.1}. A subregion with a size of
$120^{\prime\prime}$, being larger than the SUMER observation
region, was extracted from the averaged magnetogram and then used to
build a 3-D potential magnetic field, which is based on the model
proposed by \cite{Seehafer1978}. In Fig.~\ref{fig.1}, the
projections of the extrapolated magnetic loops onto the x-y-plane
have been plotted on the different images. In order to obtain a
better view of the complicated magnetic field structure, the field
lines reaching higher and lower than 40~Mm are plotted in purple and
green, respectively.

In Fig.~\ref{fig.1}, a TRACE (Transition Region and Coronal
Explorer) image \citep{Handy1999} was also shown to reveal the
chromospheric network pattern. It is an average of 100 frames
observed from 14:00 to 15:09 UT in the 1550~{\AA} passband. The
images have a pixel size of 0.5$^{\prime\prime}$. Before averaging,
the standard software for reducing TRACE data was applied to the
images, and the satellite jitter was subsequently removed by
applying the cross-correlation technique. The coalignment between
TRACE and MDI images was done through cross correlation (note that
the SOHO spacecraft was rotated at this time).

STEREO data were also available during this period. The separation angle
between the two spacecraft was 59$^\circ$ and thus allowed for a stereoscopic
study of some of the events and structures in the quiet Sun. The data obtained
from 12:00 to 17:00 UT by the two almost identical SECCHI/EUVI telescopes
\citep{Wuelser2004} in all the four passbands were reduced by using the procedure
\textit{SECCHI\_prep.pro} available in SSW (SolarSoft).

\section{Results and Discussion}
\subsection{Continuous mass supply to magnetic funnels}

The prominent blue shifts of Ne~{\sc{viii}} associated with network
junctions in the quiet Sun were first reported by \cite{Hassler1999}.
However, so far the meaning of this blue shift and its relationship with the
persistent red shift in lines at lower temperatures have not been understood.
\cite{Tian2008a} revisited the middle-latitude quiet-Sun data analyzed before
by \cite{Hassler1999}. With the help of magnetic-field extrapolation,
they found that most of the blue shifts seem to be associated with the
legs of magnetic loops, and most likely indicate plasma outflow into and
mass supply to coronal loops.

From our Fig.~\ref{fig.1} we find that almost all of the patches
that have a significant blue shift on the dopplergram of
Ne~{\sc{viii}} coincide with legs of loops located at network
junctions. Thus, we confirm the previous finding in
\cite{Tian2008a}, who also noticed that there are loops revealing
large blue shifts in both legs, and some loops with upflow in one
and downflow in the other leg. However, from Fig.~\ref{fig.1} we can
see now that most of the blue-shift patches coincide with both legs
of magnetic loops, and some patches are associated with the common
leg of several joint loops. Loop legs may generally be shaped in the
form of funnels. \cite{Peter2001} proposed a TR structure, in which
magnetic funnels can either be connected to the solar wind, or form
the legs of large coronal loops. In this picture, one funnel just
corresponds to one leg of a single loop. However, in
Fig.~\ref{fig.1} we find that a single funnel can in fact be a
common leg of several joint loops with different spatial scales and
orientations. Thus, the mass and energy flowing into a single funnel
can then be spread and supplied to multiple loops. In some cases the
flows in different loops may have different velocities and thus
reveal several sub-patches of blue shift within one blue-shift patch
on the dopplergram of Ne~{\sc{viii}}.

\cite{Marsch2008} proposed the concept of coronal circulation, or
convection to use a more apt term first coined by \cite{Foukal1978},
to emphasize that the plasma in the corona is nowhere static but
everywhere flowing, being thereby guided by various magnetic
channels. Many kinds of flow appear to be long-lasting on large
scales, and thus may indicate quasi-steady plasma convection
encompassing and affecting the entire corona and TR. In particular,
the blue shifts of Ne~{\sc{viii}} at the network junctions are
observed to be long lasting, and thus should play a permanent role
in the process of coronal mass circulation. It is interesting to
note that in the network of the quiet Sun the ultraviolet emission
lines formed in the TR are usually redshifted by a few km/s
\citep[e.g.,][]{Doschek1976,Brekke1997,Chae1998,Curdt2008}. This
phenomenon can be seen in the dopplergram of N~{\sc{iv}} as
presented in Fig.~\ref{fig.1}. Here the Doppler shift was determined
by assuming a net average shift of zero in the entire region. The
origin of this red shift is still under debate \citep[see reviews
in][]{Mariska1992,Brekke1997}. From Fig.~\ref{fig.1} it is clear
that most loop legs are associated with patches of strong
N~{\sc{iv}} red shift. However, the strongest red shifts of
N~{\sc{iv}} do not fully coincide with but slightly deviate from the
strongest blue shifts of Ne~{\sc{viii}}, which was found previously
by \cite{Aiouaz2008} and \cite{Tian2008c}.

Also, the relationship between the red shift of cool lines and blue
shift found at higher temperatures is not well understood. In
coronal holes, the contemporaneous and adjacent red and blue shifts
were explained as indicating downflow and upflow after magnetic
reconnection between open field lines in coronal funnels and their
side loops \citep{Axford1999,Tu2005a,He2008}. In the quiet Sun, the
scenario of continuous reconnection might also apply, if the
magnetic polarities of side loops are opposite to those of
funnel-like loop legs \citep{McIntosh2007,Tian2008a,Aiouaz2008},
thus enabling reconnection.

There might be another possibility: Cool plasma (in photosphere and
chromosphere) might continuously enter any loop leg through a
certain process (e.g., diffusion) from outside, but then flow up and
speed up after heating occurred. At the height of the upper TR, this
flow becomes significant and may lead to a strong blue shift of the
emission lines formed there. Due to the onset of possible
(radiative) cooling effects \citep[e.g.,][]{Kamio2009}, the flow
might again decelerate above a certain height (perhaps in the lower
corona) and finally turn downwards and accelerate under gravity,
which may lead to emission by the dense plasma at lower temperatures
and then cause the red shift of TR lines. The steadiness of the
observed shifts suggests that all these processes should occur
continuously and persistently. However, since the density decreases
with height, the contribution of the cooling plasma to the observed
red shift at TR temperatures might be minor, as compared to that of
the downward plasma resulting from reconnection.

\subsection{Relationship between mass flux and expansion factor}

As mentioned in \cite{Tian2008a}, we can estimate the rate of mass
supply to a coronal loop if we accept the Doppler shift of
Ne~{\sc{viii}} as a proxy for the plasma bulk flow (i.e., of the
proton flow). The mass flux can be calculated as $f=N_eVA$, where
$N_e$ ,$V$ and $A$ represent electron density, outflow velocity and
the area of an observed blue-shift patch. Here we selected 21
patches with significant blue shifts, which are clearly associated
with funnel-like loop legs, as inferred from Fig.~\ref{fig.1}. For
each case, we selected the region where the blue shift is larger
than 3~km/s, by plotting contours on the dopplergram, and then
calculated $A$ and the average $V$. For our studied region, we could
not make a direct density measurement by using the available data.
However, under the assumption of the same thermal structure, we can
evaluate the electron density by using the square root of
Ne~{\sc{viii}} line intensity \citep{Xia2003,Marsch2004}. Since the
Mg~{\sc{vii}} and Ne~{\sc{viii}} lines have a similar formation
temperature, we adopted a value of $N_e=10^{8.95}$cm$^{-3}$, from
the density measurement by using Mg~{\sc{vii}} in \cite{Landi1998}.

By assuming this value for the average density of our studied region, we
got a scaling factor between the electron density and the square root of the
intensity. We then superposed contours of the Doppler shift on the intensity
image and thus obtained the average intensity $I$ within each contour. By use
of the scaling relation, we got the values of the electron density. The
values of $V$, $A$, $I$, and the approximate coordinates of all cases
analysed are listed in Table~\ref{table1}. We have to mention that the
Doppler shift was determined by assuming a net average shift of zero in the
entire region. Since the average Doppler shift of Ne~{\sc{viii}} is about
2~km/s (blue shift) in the quiet Sun \citep{PeterJudge1999,Xia2004},
we simply added 2~km/s to $V$ before we calculated the mass flux for each case.

The information on the 3-D coronal magnetic field, as obtained by an
extrapolation of the measured photospheric magnetic field, allows
one to study the geometry of magnetic flux tubes in the corona.
Since plasma flows are guided by magnetic flux tubes, the expansion
of a tube might play an important role in the process of coronal
mass supply. Since the magnetic flux is conserved along a flux tube,
we can calculate the expansion factor of the tube (see, e.g.,
\cite{Marsch2004}) if the values of magnetic field strength at
different heights in the tube are known. Here we aimed at a
calculation of the expansion factor of each funnel-like loop leg
below 4~Mm. The selection of 4~Mm is based on the estimation of
Ne~{\sc{viii}} emission height in the quiet Sun \citep{Tu2005b}.

On the magnetogram shown in Fig.~\ref{fig.1}, strong-field regions
(larger than 40~gauss in magnitude) in the photosphere are outlined
in yellow. For each selected patch of blue shift on the dopplergram,
it is easy to find the associated flux tube (strong-field region) on
the magnetogram. These contours were then superposed on the
extrapolated magnetogram at 4~Mm. We calculated the average values
of the longitudinal magnetic field strength (\textit{$B_z$}) in all
contours which are associated with the selected funnels.
Table~\ref{table1} lists the results. The ratio of \textit{$B_z$} at
0~Mm and 4~Mm is a measure of the area ratio between these two
heights, the requested expansion factor.

\begin{figure}
\centering
{\includegraphics[height=0.28\textheight,width=0.5\textwidth]{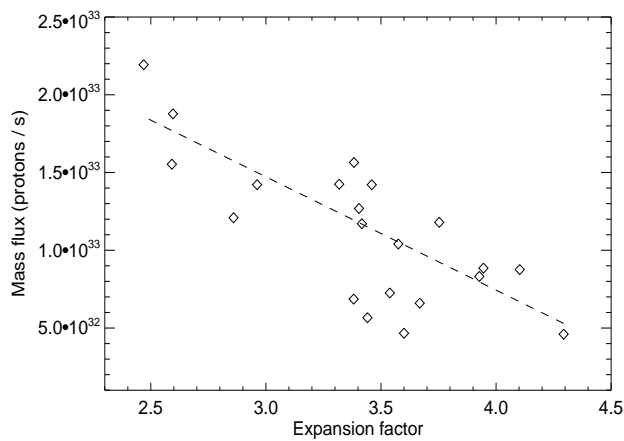}}
\caption{Scatter plot of the mass flux versus expansion factor for
different magnetic funnels. The dashed line is a linear fit to the
scattered data points. } \label{fig.2}
\end{figure}

Fig.~\ref{fig.2} presents a scatter plot of the mass flux versus
expansion factor for different magnetic funnels. The dashed line is
a linear fit to the scattered data points. We can conclude that
there is a declining trend of the mass flux with increasing
expansion factor. It is known that the source regions of fast solar
wind, the polar coronal holes, are characterized by a relatively
slow flux-tube expansion. Typical sources of the slow solar wind are
the many small holes located adjacent to active regions and the
boundaries of polar holes, where rapidly diverging fields are known
to be dominant \citep[see a review in][]{Wang2009}. The relationship
between mass flux and expansion factor here is similar to the well
known relationship between wind speed and expansion factor. Thus a
similar explanation might apply to both cases, although the spatial
scales are different. In a rapidly diverging magnetic funnel, most
of the energy that is brought in by the cool plasma and produced in
heating processes may be deposited in a lower layer of the funnel
(below the formation height of Ne~{\sc{viii}}), and therefore the
energy used to drive the upflow will be reduced. In contrast, if the
expansion of the funnel is not significant, more energy will become
available to accelerate the upflow, and thus the upward mass flux
will be increased.

\begin{table}[]
\caption[]{~Parameters derived from the dopplergram and intensity
image of Ne~{\sc{viii}}, and from the extrapolated magnetograms.
\textit{V} and \textit{I} are the average velocity and intensity in
each blue-shift patch with area \textit{A}. \textit{$B_z$}
represents the magnitude of the longitudinal component of the
extrapolated magnetic field vector.} \label{table1}
\begin{center}
\begin{tabular}{p{2cm} p{2.5cm} p{2.0cm} p{2.5cm} p{2.2cm} p{2.2cm}}
\hline\hline
Position (x$^{\prime\prime}$, y$^{\prime\prime}$)  & \textit{V}~(km/s)
& \textit{A}~(Mm$^2$) &  \textit{I}~(count rate) & \textit{$B_z$} at 0~Mm (gauss) & \textit{$B_z$} at 4~Mm (gauss) \\
\hline
-175,55 & 5.11 & 165.57 & 3.80 & 83.70 & 32.3 \\
-130,75 & 5.80 & 237.46 & 3.06 & 123.7 & 50.1 \\
-80,35 & 6.01 & 141.60 & 2.73 & 74.90 & 22.0 \\
-35,75 & 7.13 & 196.07 & 2.40 & 118.9 & 45.8 \\
-145,5 & 4.95 & 54.464 & 3.31 & 66.60 & 18.5 \\
-110,-50 & 5.39 & 156.85 & 2.26 & 91.20 & 24.3 \\
-160,-75 & 5.62 & 76.250 & 3.40 & 73.60 & 20.8 \\
-75,-145 & 4.79 & 78.428 & 3.35 & 68.60 & 18.7 \\
-45,-95 & 7.50 & 128.53 & 2.14 & 67.20 & 23.5 \\
-25,-40 & 5.05 & 115.46 & 2.58 & 64.70 & 16.4 \\
-60,-10 & 4.67 & 119.82 & 2.62 & 83.70 & 20.4 \\
55,-90 & 7.04 & 143.78 & 2.61 & 69.60 & 23.5 \\
100,-60 & 5.68 & 152.50 & 2.18 & 92.60 & 27.1 \\
115,-120 & 4.51 & 108.92 & 3.00 & 97.00 & 24.7 \\
150,-110 & 4.84 & 56.643 & 3.08 & 83.30 & 19.4 \\
145,-20 & 4.55 & 215.67 & 2.22 & 81.30 & 24.5 \\
165,90 & 6.92 & 150.32 & 2.97 & 89.30 & 26.4 \\
80,10 & 4.55 & 189.53 & 2.86 & 69.20 & 20.0 \\
135,-40 & 4.24 & 95.857 & 2.87 & 72.70 & 21.5 \\
-5,-110 & 4.03 & 91.500 & 2.30 & 58.50 & 17.0 \\
10,-130 & 4.38 & 174.28 & 1.91 & 52.20 & 14.6 \\
\hline
\end{tabular}
\end{center}
\end{table}

\subsection{Upflows in coronal bright point}

Coronal bright points (BPs) are characterized by locally enhanced
emission in X-ray, EUV and radio wavelengths and related to bipolar
magnetic fields at network boundaries
\citep{Habbal1990,Webb1993,Falconer1998,Brown2001,Madjarska2003}.
It is believed that BPs are associated with small-scale (typically
$30^{\prime\prime}-40^{\prime\prime}$ in size) loop structures
\citep{SheeleyGolub1979,Tian2007,Perez2008}.

In our field of view we found a typical BP, located and visible in
the upper left corner of all panels of Fig.~\ref{fig.1}, where we
can see a magnetic loop system with both legs being anchored in the
magnetic network. The BP shows enhanced emission in the intensity
map of Ne~{\sc{viii}}. The most interesting feature here is that
upflows are seen in both legs of the loop system (upper right
panel). Recently, \cite{Brosius2007} reported upflow and downflow on
opposite sides of a BP, which was explained as the result of
magnetic reconnection. \cite{Tian2008b} found a BP which revealed a
totally different upflow/downflow boundary at lower and higher
temperatures, suggesting a twist of the associated magnetic loop
system. According to our knowledge, this is the first time that
upflows are found in both legs of a BP loop system. This
Doppler-shift pattern is not strange, since the BP is associated
with a magnetic loop, which should not be too different from other
loop structures, although this BP-loop is of smaller size. Thus we
may conclude that the blue shift of Ne~{\sc{viii}} seen in both legs
of the BP loop system is more likely a signature of mass supply to
these loop, rather than a signature of solar wind origin. Our
conclusion is consistent with the one made earlier by
\cite{Wilhelm2000} and \cite{XiaEtal2003}, who concluded that BPs do
not directly contribute to solar wind outflow. However, other
authors suggested that BPs could be associated with jets
\citep[e.g.,][]{Shibata1996,Yokoyama1996} and might contribute to
the high-speed solar wind in coronal holes \citep{Cirtain2007}. In
the quiet Sun, if magnetic field lines are transiently open, then
their interactions with emerging flux can produce BPs and jets,
releasing plasma into the outer corona and solar wind. More studies
are needed to investigate the role of BPs in solar wind origin.

\begin{figure}
\centering {\includegraphics[width=\textwidth]{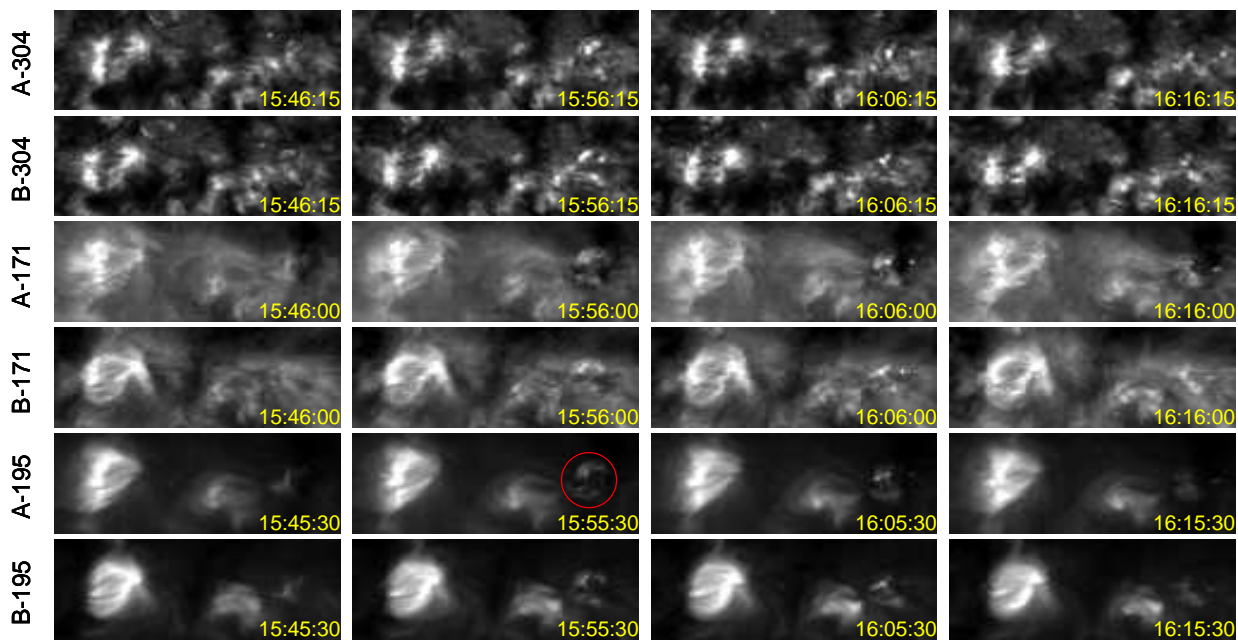}}
\caption{Time sequences of EUVI images taken in three passbands on
the two STEREO spacecraft. The observation time is shown in the
lower right corner of each image. The field of view corresponds
approximately to an x-coordinate range from $-200^{\prime\prime}$ to
$40^{\prime\prime}$ and a y-coordinate range from
$20^{\prime\prime}$ to $120^{\prime\prime}$. The red circle
indicates the position of the transient siphon flow. Movies showing
the evolution of the emissions in all four passbands of EUVI can be
found online.} \label{fig.3}
\end{figure}

The EUVI/SECCHI images presented in Fig.~\ref{fig.3} reveal the
varying morphology of the BP when seen from two different viewing
angles. The general emission patterns of the BP in the three
passbands did not change during the period of SUMER observation,
although the fine structures were rather dynamic. A comparison
between the extrapolated magnetic structure and the BP emissions as
seen by EUVI, TRACE, and SUMER suggests a remarkable agreement and
thus indicates an almost steady, potential-like nature of the
magnetic field. Our finding confirms the result of \cite{Perez2008},
who found an agreement between the extrapolated magnetic field
configuration and some of the loops composing the BP as seen in the
X-ray images and suggested that a large fraction of the magnetic
field in the BP is close to being a potential field. The remarkable
agreement here also suggests the suitability of the potential-field
model for our study.

It is also clear that the emission pattern of the BP is different at
different temperatures. At chromospheric temperatures, bright
emission can be found only at the very bottom of loop legs, as seen
in the TRACE 1550~{\AA} image. With increasing temperature, from
EUVI 304~{\AA}, SUMER N~{\sc{iv}}, SUMER Ne~{\sc{viii}}, EUVI
171~{\AA}, to EUVI 195~{\AA}, more and more parts of the upper
sections of the loop system are revealed. So it is difficult to
identify the full loop structure from only one image. By inspection
of Fig.~\ref{fig.3}, we can also find that this BP is differently
resolved in the various EUVI images onboard the two STEREO
spacecraft, suggesting the importance of stereoscopic observations.
As pointed out by \cite{Peter2007}, 3-D models are important to
account properly for the plasma and the magnetic field structure as
well as their interaction. Our results indicate, from an
observational point of view, the importance of carrying out combined
studies that use 3-D observations at different temperatures and
magnetic field extrapolation for loop-like structures. The 3-D
reconstruction of active-region loops has been successfully done in
the recent past \citep{Feng2007}. In principle, it should also be
possible to reconstruct the 3-D structure of a small-scale loop
system that, e.g., is associated with a BP.

\subsection{A transient siphon flow outside magnetic funnels}

It is well known that the corona is rather dynamic, not only in active
regions but also in the quiet Sun. Small-scale transient events have been
reported frequently for the quiet Sun. In our data set we also found one
case of a dynamic event, a transient siphon flow.

This siphon flow occurred approximately from 15:50 to 16:10~UT on
the right side of a blue-shift patch (the one with a coordinate of
(-35,75) in Table~\ref{table1}). It was recorded in all of the four
passbands of EUVI on both spacecraft. Movies showing the evolution
of the emissions in the four passbands can be found online (note the
different cadences in different passbands). We did not de-rotate the
images since an interpolation might have smoothed out the
small-scale event. Fig.~\ref{fig.3} only shows several snapshots.
From the movie we can see bright emission features moving along two
small parallel loops from one end to the other. The two loops were
located in a weak mixed-polarity region on the magnetogram shown in
Fig.~\ref{fig.1}. They were rather small and cold before the
occurring of the event, and thus were not properly resolved by EUVI.
Due to the elevated temperature and enhanced coronal emission
resulting from an unknown heating process, the two cold loops were
clearly seen during the period of the siphon flow, especially at
around 15:55:30 in the 195~{\AA} passband of EUVI onboard STEREO-A.
Under the assumption of a semi-circular shape, the loop length can
be estimated as 40$^{\prime\prime}$. Then the speed of the flow can
be calculated as 24~km/s.

Siphon flows are believed to be driven by asymmetric heating or
pressure gradients between two legs of the loop
\citep[e.g.,][]{Robb1992,Orlando1995}. They have been frequently
found in large active-region loops, but rarely been observed in the
quiet Sun. \cite{Teriaca2004} identified a supersonic siphon-like
flow in a quiet-Sun loop, by analyzing spectral profiles of
O~{\sc{vi}}. Siphon flows in small loops in active regions were also
detected by \cite{Uitenbroek2006} and \cite{Doyle2006}. Here we
reported a direct observation of a small-scale siphon flow in the
quiet Sun by both STEREO satellites. However, since this transient
siphon flow occurred in a weak mixed-polarity field region and was
outside the adjacent magnetic funnel, it should not be related to
the long-lasting upflow in the funnel.

\subsection{Flows in the magnetic structures of the quiet Sun}

\begin{figure}
\centering {\includegraphics[width=\textwidth]{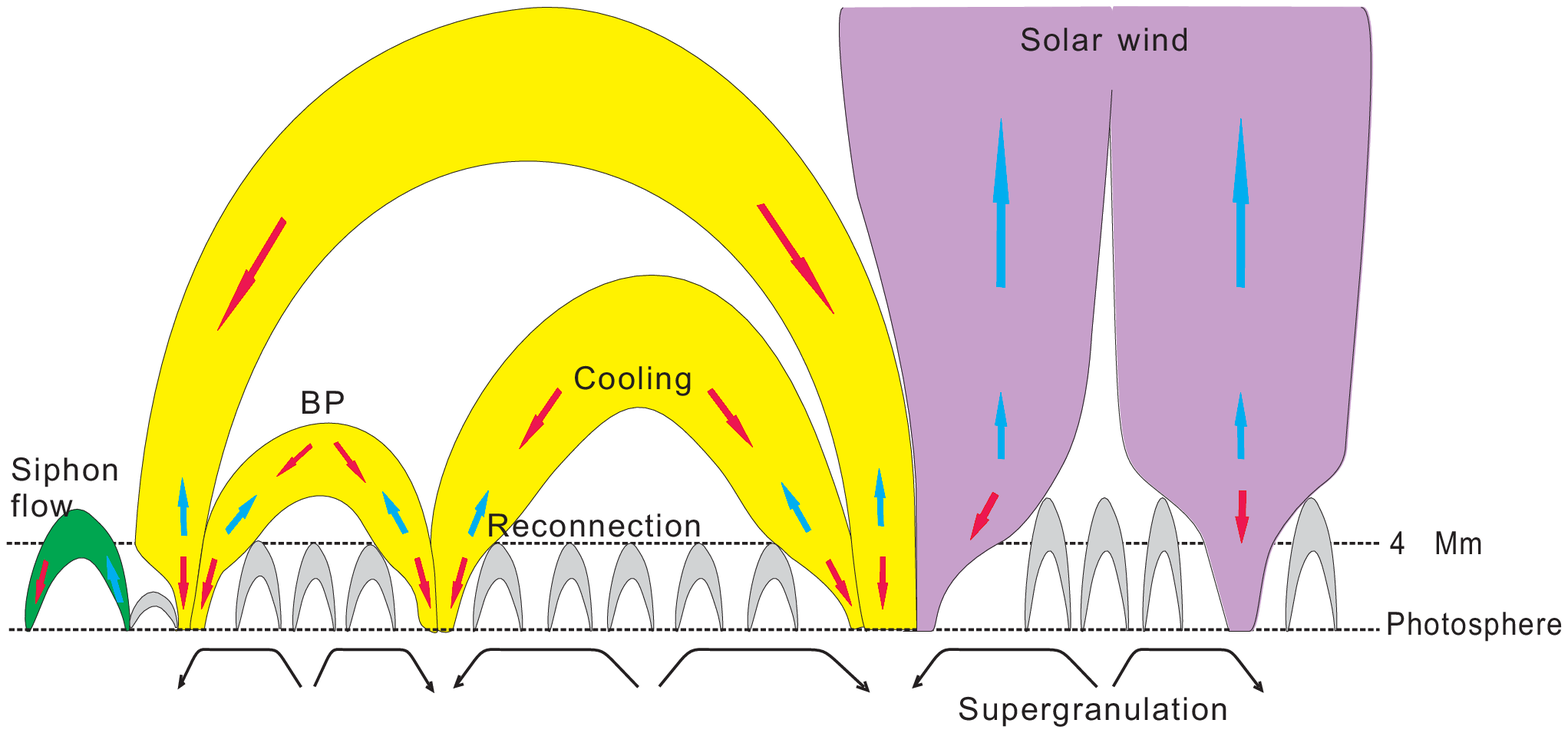}}
\caption{Schematic presentation of the flows in the magnetic
structures of the quiet Sun. Magnetic loops with different scales
and open-field regions are marked in different colors. Downflows and
upflows are indicated by the red and blue arrows, respectively. }
\label{fig.4}
\end{figure}

Based on the above observations, we suggest a quiet-Sun magnetic
structure which is presented in Fig.~\ref{fig.4}. The cool loops,
coronal loops, and open-field structures are marked in grey, yellow,
and purple, respectively. Downflows and upflows guided by these
structures are indicated by the red and blue arrows, respectively.

Cool loops with different spatial scales can be found in both
network and internetwork regions. Legs of one or several large-scale
magnetic structures (coronal loops or open-field structures) may be
anchored in and crowd a network junction, forming a magnetic funnel.
The internetwork loops are continuously swept through the
supergranular convection to the network boundaries, where they may
interact with the preexisting funnels. These funnels can either be
connected to the solar wind, or form the legs of large coronal
loops.

As mentioned previously, two mechanisms might be responsible for the
systematic flows in coronal loops. Continuous reconnection between
field lines in magnetic funnels and side loops is expected to
produce upflows in the upper TR and downflows in the lower TR. The
outflows produced by reconnections around a funnel tend to converge
towards the center of the funnel. In contrast, the hot plasma
trapped in low-lying loops are pulled down when they cool, and the
downflows are stronger at the boundary of the network where side
loops are accumulated. Thus, the bi-directional flows are likely to
be detected as the not-fully-cospatial blue shift of Ne~{\sc{viii}}
and red shift of N~{\sc{iv}}. On the other hand, heating and cooling
processes might also take place in quiet-Sun coronal loops. In most
cases, upflows which are possibly caused by heating can not reach
the apices of the loops. When cooling is switched on, the flows turn
downwards and lead to emission at lower temperatures. This scenario
will naturally explain the observational fact that most of the
strongest red shifts of N~{\sc{iv}} do not fully coincide with the
strongest blue shifts of Ne~{\sc{viii}}, since the turning points
are located in the curved segments of loop legs.

In classical pictures \citep[e.g.,][]{Peter2001}, the dominant type
of flow in coronal loops is siphon flow. However, our observation
seems to indicate that siphon flow rarely exists in quiet-Sun
coronal loops with a size comparable to or larger than a
supergranule. In these large-scale loops, the upflowing plasma will
get cooled and turn downwards before reaching the apices. However,
transient heating in one leg of a small loop may launch an upflow
that can reach the apex and subsequently flow downwards along the
other leg of the loop. For BPs which are associated with magnetic
loops of intermediate scales these two types of flows may both
exist.

The mass supplied into the coronal loops might also be released into
the ambient corona or even into the solar wind. As claimed by
\cite{He2007} and \cite{Tian2008a}, sometimes quiet-Sun coronal
loops might transiently open due to magnetic reconnection so that
open field lines forming magnetic funnels might also be present in
the quiet Sun. These funnels originating from different network
regions expand with height and finally merge into a single wide
open-field region. The nascent solar wind is likely to be produced
in these funnels, but at locations higher than the source of the
Ne~{\sc{viii}} emission. Recent EIS observations showed that coronal
lines revealed high outflow velocities on the order of 100~km/s in a
compact region (network boundary) of the quiet Sun \citep{Dere2007}.
Future combined studies between magnetic field extrapolation and
spectroscopic observations including SUMER and EIS are needed to
investigate whether this high-speed coronal outflow corresponds to
the quiet-Sun solar wind or not.

\subsection{Implications for coronal heating and unresolved magnetic structures}

\begin{figure}
\centering {\includegraphics[width=0.5\textwidth]{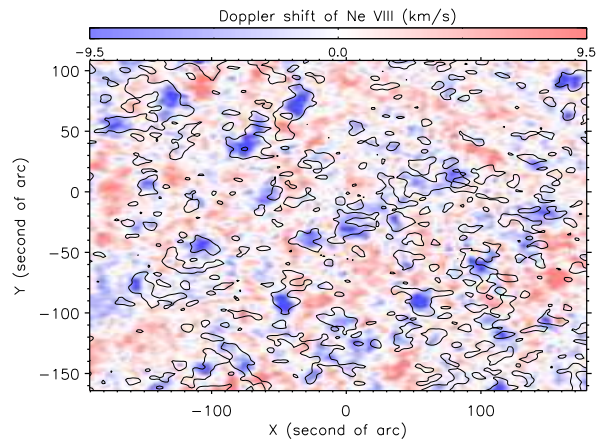}}
\caption{Contours of line width (top 20\%) superposed on the
dopplergram of Ne~{\sc{viii}}. } \label{fig.5}
\end{figure}

EIS observations reveal hot plasma upflows of several tens km/s and
enhanced nonthermal velocities near the footpoints of active region
loops \citep{Hara2008}. The authors claimed that this result
supports the nanoflare heating model of \cite{Patsourakos2006} that
treats a coronal loop as a collection of unresolved small-scale
bundles. While based on a more recent finding that SUMER and EIS
observations seem to show faint upflows at 50-100~km/s for
temperatures from 100,000 to several million degrees,
\cite{DePontieu2009} suggested that the dominant part of coronal
heating is provided by the chromospheric jets or type II spicules.

A heating process is usually associated with high-speed laminar
flows, waves, or turbulent flows which all contribute to the
nonthermal motion and tend to broaden the width of spectral lines
\citep[e.g.,][]{Hollweg1984,Chae1998b,Tu1998}. Thus, an
investigation of the line width may provide some implications for
coronal heating. The correlation between the line radiance and
non-thermal width for a certain TR line was investigated by
\cite{Chae1998b}. Here we found a similar result: the correlation
coefficient is much higher for the typical TR line N~{\sc{iv}}
(0.33) than for the upper TR line Ne~{\sc{viii}} (0.13). Our most
interesting result is that there is a high correlation between the
line width and the Doppler shift of Ne~{\sc{viii}} (correlation
coefficient: -0.37, note that the blue and red shifts have negative
and positive values, respectively). Contours of line width (top
20\%) are superposed on the dopplergram of Ne~{\sc{viii}} in
Fig.~\ref{fig.5}. It is very clear that most of the significant blue
shifts are associated with an enhancement of the line width.

\cite{Chae1998b} claimed that the superposition of different laminar
flows in various unresolved loops can not explain some observational
characteristics of nonthermal motions. However, some model
calculations led to a different conclusion. Multi-thread loop models
suggested that plasma from a single thread (strand) can have a very
high speed, but since the initial rise of each thread is masked by
threads that were heated previously and are emitting strongly, the
composite emission from many strands in various stages of heating
and cooling is dominated by the stationary emission
\citep{Warren2005,Patsourakos2006}. According to this scenario, the
plasma with a long-lasting blue shift of the order of 6~km/s in our
observations should correspond to the stationary thermal emission in
the upper TR of network junctions, while emission from some
individual unresolved threads with high-speed motions corresponds to
heating beams which yield increments of the line width but no
detectable upflows at upper-TR temperatures. These high-speed upward
plasma may be associated with the chromospheric jets or type II
spicules, which were suggested to be crucial for coronal heating in
the mechanism proposed by \cite{DePontieu2009}. In our observation,
most Ne~{\sc{viii}} profiles are well approximated by a single
Gaussian function and do not show visible asymmetries, which is
consistent with previous observations and the nanoflare heating
model of \cite{Patsourakos2006}.

However, we have to keep in mind that the enhancement of the line
width in network junctions may also be related with waves or
turbulence, which are associated with the process producing the
significant Ne~{\sc{viii}} blue shift.

\section{Summary}

Through a potential-field extrapolation, we have further
investigated the coupling between solar magnetic structures and the
prominent blue shifts of Ne~{\sc{viii}} at the network junctions in
the quiet Sun. An exceptionally clear and close relationship between
the significant upflows and the funnel-like loop legs has been
established. Moreover, we have made a first attempt to study the
influence of the expansion factor of magnetic loops on the mass flux
as inferred from spectroscopic observation and found an
anti-correlation.

For the small-scale loop structure identified as coronal bright point, we
have made the first combined study between the extrapolated magnetic field
and 3-D multi-temperature EUV observations. We find upflows in both
legs of the loop system associated with the BP.

We also reported a direct detection of a transient siphon flow in
the quiet Sun by both STEREO satellites. This siphon flow along two
parallel small-scale loops occurred in a mixed-polarity-field
region, which was located outside the adjacent magnetic funnel.

Based on these observations, we present in Fig.~\ref{fig.4} a new
scenario of the flows guided by the quiet-Sun magnetic structures.
In this scenario, the dominant flows at upper-TR temperatures in
quiet-Sun coronal loops are long-lasting upflows rather than siphon
flows. The mass supplied into and flows upwards along coronal loops
may fall downwards again when cooling is switched on. Sometimes the
coronal loops might transiently open due to magnetic reconnection
and thus can release mass into the ambient corona or even into the
solar wind.

Finally, we have recalled some recent results on the upflows in the
TR and corona, and discussed the implications of our results for
coronal heating. The significant correlation between the line width
and Doppler shift of Ne~{\sc{viii}} seems to favor a scenario of
magnetic funnel consisting of unresolved strands.

\begin{acknowledgements}
{\bf Acknowledgements:} The SUMER project is financially supported
by DLR, CNES, NASA, and the ESA PRODEX programme (Swiss
contribution). SUMER and MDI are instruments on board SOHO, an ESA
and NASA mission. TRACE and STEREO are NASA missions. The
SECCHI/STEREO data used here were produced by an international
consortium of the Naval Research Laboratory (USA), Lockheed Martin
Solar and Astrophysics Lab (USA), NASA Goddard Space Flight Center
(USA), Rutherford Appleton Laboratory (UK), University of Birmingham
(UK), Max-Planck-Institut for Solar System Research (Germany),
Centre Spatiale de Li\`{e}ge (Belgium), Institut d'Optique
Th\'{e}orique et Applique\'{e} (France), and Institut
d'Astrophysique Spatiale (France). Hui Tian thanks Dr. Suguru Kamio
for the helpful discussion on the Hinode observation. Hui Tian is
supported by the IMPRS graduate school run jointly by the Max Planck
Society and the Universities of G\"ottingen and Braunschweig. The
work of Hui Tian's group at Peking University is supported by the
National Natural Science Foundation of China (NSFC) under contract
40874090.
\end{acknowledgements}

\end{document}